# Dispersive effects during long wave run-up on a plane beach

Ahmed Abdalazeez[1], Ira Didenkulova[1,2*] and Denys Dutykh[3]

[1] Department of Marine Systems, Tallinn University of Technology, Tallinn, Estonia
[2] Nizhny Novgorod Technical University n.a. R.E. Alekseev, Nizhny Novgorod, Russia
[3] Univ. Grenoble Alpes, Univ. Savoie Mont Blanc, CNRS, LAMA, 73000 Chambéry, France
*corresponding author: irina.didenkulova@taltech.ee

**Abstract.** Dispersive effects during long wave run-up on a plane beach are studied. We take an advantage of experimental data collection of different wave types (single pulses, sinusoidal waves, bi-harmonic waves, and frequency modulated wave trains) and simulate their run-up using two models: (i) non-dispersive nonlinear shallow water theory and (ii) dispersive Boussinesq type model based on the modified Peregrine system. It is shown, that for long positive pulses, dispersive effects are not so important and nonlinear shallow water theory can be used. However, for periodic sinusoidal and bi-harmonic pulses of the same period, the dispersive effects result in significant wave transformation during its propagation, but do not have a strong impact on its maximal run-up height. Overall, for maximum wave run-up height, we could not find a preference of dispersive model against the nondispersive one, and, therefore, suggest using nonlinear shallow water model for long wave run-up height estimation.

## 1 Introduction

There are several reasons why the nonlinear shallow water theory (NLSW) is favored for long wave run-up calculations as compared to dispersive wave models, often represented by Boussinesq-type approximations. First of all, wave run-up calculated using dispersive codes is prone to numerical instabilities, which make computations more sensitive to numerical parameters (Bellotti and Brocchini, 2002). Second, the Boussinesq terms in dispersive model tend to zero at the shoreline, so that dispersive equations simplify to NLSW in this region (Madsen et al. 1997).

Horrillo et al. (2006) studied dispersive effects during 2004 Indian Ocean tsunami propagation by comparing NLSW with the fully nonlinear Navier-Stokes equations (FNS). They came to conclusion that, NLSW is more suitable in hazard assessments (e.g. it has a very low computation cost and often it over-predicts the maximum wave run-up height, whereas this property is consider a safety factor). Moreover, Glimsdal et al. (2013) suggested that NLSW is more appropriate for warning purposes. However, dispersive effects become more important in trailing waves, whereas the leading waves can be well described by NLSW (Løvholt et al. 2014). Note, that maximum wave is often not the first one, at least for tsunamis propagating over a long distance, see, for example, Candella et al. (2008).

Most of the mentioned studies were based on numerical results and were missing the fidelity control mechanism. Therefore, in this paper we take an advantage of available experimental data collection of different wave types (single pulses, sinusoidal waves, bi-harmonic waves, and frequency modulated wave trains) and simulate their run-up using two models: (i) NLSW, and (ii) dispersive model of Boussinesq type based on the modified Peregrine system (mPer).

The paper is organized as follows. In Section 2, we describe the available experimental dataset. The numerical models (NLSW and mPer) are briefly described in Section 3. In Section 4, we compare numerical results of NLSW and mPer models with the experimental data. The main results are summarized in Section 5.

## 2 Experimental data

The experiments were conducted in the Large Wave Flume (GWK), Hannover, Germany. The experimental set-up consisted of a 251 m long section of constant depth 3.5 m and a plane beach with a slope angle 1:6 (fixed asphalt bed). There were from 16 to 18 wave gauges recording wave propagation along the flume. The wave run-up was measured by capacitance probe, which was supplemented by two regular video cameras. Waves were generated by the piston type wave maker, which showed to be efficient for long wave generation (Schimmels et al. 2016). The generated waves are listed in Table 1. Details of this experiment can be found in Didenkulova et al. (2013).

**Table 1.** Generated waves and their run-up heights

| Type of waves | Wave period (s) | Initial wave amplitude (m) | Experimental run-up (m) | NLSW run-up (m) | mPer run-up (m) |
|---|---|---|---|---|---|
| Positive pulse | 20 | 0.10 | 0.259 | 0.268 | 0.254 |
| Positive pulse | 20 | 0.24 | 0.795 | 0.840 | 0.780 |
| Sine wave | 20 | 0.05 | 0.096 | 0.13 | 0.14 |
| Sine wave | 20 | 0.60 | 2.27 | 1.80 | 1.81 |
| Bi-harmonic wave | 20 | 0.12 | 0.79 | 0.89 | 0.85 |
| Bi-harmonic wave | 20 | 0.15 | 1.3 | 1.37 | 1.32 |
| Wake-like train | 20→10 | 0.10 | 0.46 | 0.60 | 0.51 |
| Wake-like train | 20→10 | 0.40 | 2.14 | 1.68 | 2.57 |

## 3 Numerical set-up

In the present work, two different numerical models are used: the non-dispersive non-linear shallow water model (NLSW) and dispersive Bousinessq type model based on the modified Peregrine system (mPer). Both models were set not taking into account bottom friction and assumed the fluid being perfect and the flow to be incompressible and irrotational.

The model left boundary conditions corresponded to the experimental wave conditions in Table 1. On the right, we placed a sufficiently long plane beach to avoid any interactions with the right boundary.

More precisely, the bathymetry in numerical experiments was set up to reproduce the Large Wave Flume conditions:

$$h(x) = \begin{cases} h_0, & x \in [a,b] \\ h_0 - (x-b)\tan\alpha, & x \in [b,c] \end{cases}, \tag{1}$$

where $h_0 = 3.5$ m is the constant water depth, $\alpha$ is the bottom slope ($\tan\alpha = 1:6$). The distance $x \in [a,b]$ is the cell interfaces, $x_c$ is the centre of a cell, $[a, c]$ are left and right boundaries of the numerical flume ($a = 0$ m), and $b = 251$ m is the point where the slope starts. The distance $x \in [a,b]$ has been divided into number of cells $c_i = (X_{i-1/2}, X_{i+1/2})$, where $X_i$ is the center of cell $i$, $X_i = 1/2(X_{i-1/2} + X_{i+1/2})$.

The numerical time steps can be calculated as $\Delta t = \left(\dfrac{t_f - t_i}{M}\right)$, where $t_i$ is the initial time, $t_f$ is the final time and $M$ is the number of time steps.

### 3.1 Nonlinear shallow water (NLSW) model

The 1D nonlinear shallow water equations are:

$$H_t + (Hu)_x = 0, \tag{2}$$

$$(Hu)_t + \left(Hu^2 + \frac{g}{2}H^2\right)_x = gHh_x, \tag{3}$$

where $H = h + \eta$ is the total water depth, $\eta(x, t)$ is the water elevation with respect to the still water level, $u(x,t)$ is the depth-averaged flow velocity, $h(x)$ is an unperturbed water depth described by Eq. (1), $g$ is the gravitational acceleration, $x$ is the coordinate directed onshore, and $t$ is time. We use the finite volumes method. The numerical scheme is based on the second order UNO2 reconstruction, for more details see (Dutykh et al. 2011).

### 3.2 Modified Peregrine (mPer) model

The Boussinesq equations for long dispersive wave propagation, derived by Peregrine (1967), are:

$$\eta_t + ((h+\eta)u)_x = 0, \tag{4}$$

$$u_t + u u_x + g \eta_x - \frac{h}{2}(hu)_{xxt} + \frac{h^2}{6}(u_{xxt}) = 0 \, . \tag{5}$$

This classical Peregrine system was modified by Durán et al. (2018) in order to recover the conservative form of equations.

Eq. (4) of the mass conservation in new variables becomes:

$$H_t + Q_x = 0 \, , \tag{6}$$

where $Q$ is the horizontal momentum, $H$ is the total water depth.

The momentum conservation equation obtained from Eq. (5) becomes (Durán et al. 2018):

$$\left(1 + \frac{1}{3}H_x^2 - \frac{1}{6}HH_{xx}\right)Q_t - \frac{1}{3}H^2 Q_{xxt} - \frac{1}{3}HH_x Q_{xt} + \left(\frac{Q^2}{H} + \frac{g}{2}H^2\right)_x = gHh_x \, . \tag{7}$$

Eqs. (6) and (7) are called the modified Peregrine equations and are studied in detail in Durán et al. (2018).

## 4    Results

The two models described above have been used to reproduce the Large Wave Flume experiments listed in Table 1. The water surface elevation has been recorded by different wave gauges at different distances from the wave maker, including wave run-up. The comparison simulated by NLSW and mPer models against measured experiments. The maximum wave run-up heights are shown in Table 1. For periodic waves, the comparison is made for experimentally measured wave with the maximum run-up height. The deviations of computational run-up heights from the measured ones are shown in Table 2.

**Table 2.** Deviations of computational run-up heights from the measured ones in %

| Type of waves | Initial wave amplitude (m) | Experimental run-up (m) | NLSW (%) | mPer (%) |
|---|---|---|---|---|
| Positive pulse | 0.10 | 0.259 | 3 | -2 |
| Positive pulse | 0.24 | 0.795 | 6 | -2 |
| Sine wave | 0.05 | 0.096 | 35 | 45 |
| Sine wave | 0.60 | 2.27 | -21 | -20 |
| Bi-harmonic wave | 0.12 | 0.79 | 13 | 8 |
| Bi-harmonic wave | 0.15 | 1.3 | 5 | 2 |
| Wake-like train | 0.10 | 0.46 | 30 | 11 |
| Wake-like train | 0.40 | 2.14 | -22 | 20 |

It can be noticed that for long positive pulses results of both NLSW and mPer models are in a good agreement with experimental data (Fig. 1). In the weak amplitude case NLSW overestimates the experimental result by 3%, while mPer underestimates it by 2%. For the initial wave of higher amplitude, the discrepancy is also higher: 6% for NLSW and 2% for mPer (Table 2). Therefore, one can conclude that the dispersive effects are not so important to predict the run-up height for the class of long single waves of positive polarity and the NLSW can be used.

For dispersive wake-like trains and even for periodic waves (sine waves and bi-harmonic waves) the situation is quite different. Most of the results of both NLSW and mPer have an accuracy of 20-30 % and do not really show the preference of one model against the other. Note, most of wave gauges located along the basin demonstrate the importance of dispersive effects, which are captured by mPer model and ignored by NLSW, but according to Table 2 they do not have principal impact on maximum run-up height. This somehow supports the existing general opinion that for the maximum wave run-up estimation NLSW is enough.

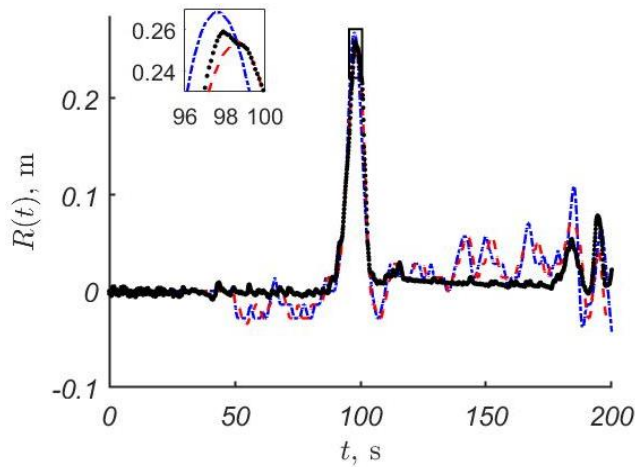

**Fig. 1**. Run-up height time-series for a positive pulse of $A$ = 0.1 m, $T$ = 20 s, calculated by mPer (red dashed line), NLSW (blue dash-dotted line) and measured experimentally (black dots).

## 5    Concluding Remarks

We used the nonlinear shallow water model (NLSW) and dispersive Boussinesq type model based on the modified Peregrine system (mPer) to reproduce physical experiments of long wave propagation and run-up. The studied waves were of different shapes and amplitudes and included long single pulses of positive polarity, periodic sine and bi-harmonic waves and dispersive wake-like wave trains.

It is found, that in our wave collection, the single pulses had the best agreement with both models and dispersive effects in this case were negligible. For periodic waves and dispersive wave trains the dispersive effects were important, and mPer model gave much better fit to the records of tide-gauges, located along the flume. However, for maximum run-up height this better capture of dispersive effects did not play a big role, and both models resulted in 20-30 % deviation from the experimental data.

## Acknowledgements

Numerical simulation of wake-like wave trains was carried out with a financial support from Russian Science Foundation grant 16-17-00041, simulation of other wave types was supported by ETAG grant PUT1378. The experimental data were obtained within Hydralab IV Grant HyIV-FZK-03. Authors also thank the PHC PARROT project No 37456YM, which funded the authors' visits to France and Estonia and allowed this collaboration.